\documentclass[twocolumn,prl,showpacs]{revtex4}
\usepackage{graphicx}
\usepackage{dcolumn}
\newcommand{\bm}[1]{\mbox{\boldmath$#1$}}
\begin{document}
\title{Order Parameter to Characterize Valence-Bond-Solid States in
Quantum Spin Chains}
\date{\today}
\author{Masaaki Nakamura$^{1}$ and Synge Todo$^{2,3}$}
\affiliation{
 $^1$Department of Applied Physics, Faculty of Science, Science
 University of Tokyo, Kagurazaka, Shinjuku-ku, Tokyo 162-8601, Japan\\
 $^2$Theoretische Physik, Eidgen\"ossische Technische Hochschule,
 CH-8093 Z\"urich, Switzerland\\ $^3$Institute for Solid State Physics,
 University of Tokyo, Kashiwa 277-8581, Japan}
\begin{abstract}
 We propose an order parameter to characterize valence-bond-solid (VBS)
 states in quantum spin chains, given by the ground-state expectation
 value of a unitary operator appearing in the Lieb-Schultz-Mattis
 argument. We show that the order parameter changes the sign according
 to the number of valence bonds (broken valence bonds) at the boundary
 for periodic (open) systems. This allows us to determine the phase
 transition point in between different VBS states. We demonstrate this
 theory in the successive dimerization transitions of the
 bond-alternating Heisenberg chains, using the quantum Monte Carlo
 method.
\end{abstract}
\maketitle

About two decades ago, Haldane made a striking conjecture that the
integer-spin $S$ antiferromagnetic Heisenberg chain has a unique
disordered ground state with an energy gap, while for a half-odd-integer
spin there is no energy gap and the system belongs to the same
universality class as the $S=1/2$ case \cite{Haldane}.  For $S=1$, this
conjecture was confirmed by many numerical \cite{Nightingale-B} and
experimental studies \cite{Buyers}.
Affleck, Kennedy, Lieb, and Tasaki
studied an $S=1$ isotropic spin chain with special biquadratic
interactions which has the exact ground state with a finite gap
\cite{Affleck-K-L-T}. They also proposed the valence-bond-solid (VBS)
state for the Haldane gap systems, and concluded that the ground state
of the Heisenberg chain is described approximately by the VBS state.
Since the spin configurations of the VBS state show the hidden
antiferromagnetic order, den Nijs and Rommelse proposed the string order
parameter to characterize the $S=1$ Haldane phase \cite{Nijs-R},
\begin{equation}
 {\cal{O}}^{\alpha}_{\rm string} = -\lim_{|k-l|\rightarrow\infty}
  \langle\Psi_0| S^{\alpha}_{k}
  \exp\left[{\rm i} \pi\sum_{j=k+1}^{l-1} S^{\alpha}_{j}\right]
  S^{\alpha}_{l}|\Psi_0\rangle,
  \label{eqn:string}
\end{equation}
where $\alpha= x,y,z$ and $|\Psi_0\rangle$ means the ground state. Thus
this order parameter enables us to detect the VBS state {\it
indirectly}. The string order parameter was generalized to $S>1$ cases
by Oshikawa \cite{Oshikawa}.

On the other hand, Haldane's prediction was also discussed by the
Lieb-Schultz-Mattis (LSM) type argument \cite{Lieb-S-M}.  Using this
technique, Affleck and Lieb examined Haldane's conjecture
\cite{Affleck-L}.  Later, Oshikawa, Yamanaka, and Affleck generalized
this argument, and obtained a necessary condition for a gapped state
\cite{Oshikawa-Y-A}.  However, the relation between the LSM argument and
the VBS picture including the string order parameter has not been fully
understood.  In this Letter, we discuss this relation, and show that an
overlap integral appearing in the LSM argument [Eq.~(\ref{eqn:def_z})
below] plays a role of an order parameter which detects VBS ground
states {\it directly}.  We also show that it can also be applied to
determine phase boundaries in between different VBS states.  We
demonstrate this idea by the quantum Monte Carlo (QMC) simulation for
successive dimerization transitions in the bond-alternating Heisenberg
spin chains.

First, let us review the LSM argument briefly based on
Ref.~\cite{Oshikawa-Y-A} and introduce the order parameter. We consider
a periodic spin chain of length $L$ with short range interactions.  We
assume the translational (${\cal T}: S_j^{\alpha}\rightarrow
S_{j+1}^{\alpha}$) and the parity (${\cal P}:
S_j^{\alpha}\rightarrow S_{L+1-j}^{\alpha}$) invariance.  Now we
introduce the following ``twist operator''
\begin{equation}
 U\equiv\exp\left[{\rm i} \frac{2\pi}{L}\sum_{j=1}^L j S_j^z\right].
  \label{eqn:twist_op}
\end{equation}
Since this operator rotates all the spins about the $z$-axis with
relative rotation angle between the neighboring spins $2\pi/L$, it
creates spin-wave-like excitations.  Applying the twist operator $q$
times to the unique normalized ground state $|\Psi_0\rangle$ generates a
set of low-lying excited states $\{|\Psi_k\rangle\equiv
U^k|\Psi_0\rangle\}$ ($k=1,\cdots,q$).  The excitation energy of the
state $|\Psi_1\rangle$ is evaluated as $\Delta E\sim{\cal O}(L^{-1})$.
Although $|\Psi_1\rangle$ is not necessarily an eigenstate of the
Hamiltonian, there exists at least one eigenstate with energy of ${\cal
O}(L^{-1})$, if the state $|\Psi_1\rangle$ is orthogonal to the ground
state $|\Psi_0\rangle$.

In order to consider the orthogonality of these states, we introduce the
following overlap integral which plays a central role in this Letter:
\begin{equation}
 z_L^{(q)}
  \equiv\langle\Psi_0|\Psi_q\rangle
  =\langle\Psi_0|U^q|\Psi_0\rangle.
  \label{eqn:def_z}
\end{equation}
The invariance under transformations ${\cal T}$ and ${\cal P}$ yields
\begin{eqnarray}
 z_L^{(q)}
  &=&\langle\Psi_0|{\cal T}U^q{\cal T}^{-1}|\Psi_0\rangle
  ={\rm e}^{{\rm i} 2q(S_1^z-m)\pi}z_L^{(q)},
  \label{eqn:z_L1}\\
 &=&\langle\Psi_0|{\cal P}U^q{\cal P}|\Psi_0\rangle
  ={\rm e}^{{\rm i} 2qm\pi}[z_L^{(q)}]^*,
  \label{eqn:z_L2}
\end{eqnarray}
where $m$ is the magnetization per site.  Eq.~(\ref{eqn:z_L1}) shows
that $z_{\infty}^{(1)}\neq0$ is possible only when $S-m=$ integer.  In
this case, the system has a gap without breaking the translational
symmetry.  On the other hand, for a rational value $S-m=p/q$ with $p$
being an integer, $|\Psi_0\rangle$, $|\Psi_1\rangle$,$\cdots$,
$|\Psi_{q-1}\rangle$ are mutually orthogonal. This means that the system
in the $L\rightarrow\infty$ limit is gapless ($z_{\infty}^{(q)}=0$),
otherwise gapped with $q$-fold degenerate ground state
($z_{\infty}^{(q)}\neq 0$) due to the spontaneous breaking of the
translational symmetry.  Thus $q(S-m)=$ integer is a necessary condition
for a gapped state \cite{Oshikawa-Y-A}.  In this case, $z_L^{(q)}$ is
real or pure imaginary due to Eq.~(\ref{eqn:z_L2}).

\begin{figure}[t]
 \includegraphics[width=3.2in]{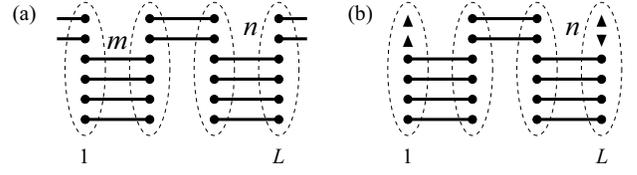}
 \vspace*{-1.0em}
 \caption{Generalized VBS states with $(m,n)=(4,2)$ for (a) a periodic
 chain and for (b) an open chain with even $L$.  Broken ovals mean the
 symmetrization of $S=1/2$ variables at each site, and solid triangles
 denote isolated bosons.}\label{fig:VBSpics}
\end{figure}

In the previous works based on the LSM argument
\cite{Lieb-S-M,Affleck-L,Oshikawa-Y-A}, however, explicit values of
$z_L$ have not been calculated.  Now let us evaluate $z_L$ for various
ground states.  In the N\'{e}el state with two-fold degeneracy, one can
obtain $z_L^{(2)}=(-1)^{2S}$ immediately.  In order to calculate $z_L$
in VBS states, we introduce the Schwinger boson representation for the
spin operators \cite{Arovas-A-H}:
\begin{equation}
 \begin{array}{cc}
  S_j^{+}=a_j^{\dag}b_j,
   &S_j^{z}=\frac{1}{2}(a_j^{\dag}a_j-b_j^{\dag}b_j),\\
  S_j^{-}=b_j^{\dag}a_j,
   &S_j=\frac{1}{2}(a_j^{\dag}a_j+b_j^{\dag}b_j),
 \end{array}
\end{equation}
where these bosons satisfy the commutation relation
$[a_i,a_j^{\dag}]=[b_i,b_j^{\dag}]=\delta_{ij}$ with all the other
commutations vanishing.  The operator $a_j^{\dag}$ ($b_j^{\dag}$)
increases the number of up (down) $S=1/2$ variables under
symmetrization.  Then, a generalized VBS state discussed in
Ref.~\cite{Guo-K-M} in a periodic system (see Fig.~\ref{fig:VBSpics}(a))
is written as
\begin{equation}
 |\Psi_{\rm VBS}^{(m,n)}\rangle
  \equiv\frac{1}{\sqrt{\cal N}}\prod_{k=1}^{L/2}
  (B_{2k-1,2k}^{\dag})^{m}(B_{2k,2k+1}^{\dag})^{n}|{\rm vac}\rangle,
  \label{eqn:VBS}
\end{equation}
where $B_{i,j}^{\dag}\equiv
a_{i}^{\dag}b_{j}^{\dag}-b_{i}^{\dag}a_{j}^{\dag}$, ${\cal N}$ is a
normalization factor, and $|{\rm vac}\rangle$ is the vacuum with respect
to bosons.  The integers $m$ and $n$ satisfy $m+n=2S$.  Using relations
$Ua_j^{\dag}U^{-1}=a_j^{\dag}{\rm e}^{+{\rm i}\pi j/L}$ and
$Ub_j^{\dag}U^{-1}=b_j^{\dag}{\rm e}^{-{\rm i}\pi j/L}$, a twisted
valence bond $UB_{j,j+1}^{\dag}U^{-1}$ for $1\leq j\leq L-1$, and that
located at the boundary are calculated as follows,
\begin{eqnarray}
 UB_{j,j+1}^{\dag}U^{-1}&=&
  {\rm e}^{-{\rm i}\pi/L}a_{j}^{\dag}b_{j+1}^{\dag}
  -{\rm e}^{ {\rm i}\pi/L}b_{j}^{\dag}a_{j+1}^{\dag},
  \label{eqn:twbond1}\\
 UB_{L,1}^{\dag}U^{-1}&=&
  -({\rm e}^{-{\rm i}\pi/L}a_{L}^{\dag}b_{1}^{\dag}
  -{\rm e}^{ {\rm i}\pi/L}b_{L}^{\dag}a_{1}^{\dag}).
  \label{eqn:twbond2}
\end{eqnarray}
In the latter case, a negative sign appears for each valence bond. This
reflects a property of an $S=1/2$ spin which needs $4\pi$ rotation to
return to the original state. Thus the asymptotic form of $z_L$ is given
by
\begin{equation}
 z_L^{(1)}=
  \langle\Psi_{\rm VBS}^{(m,n)}|U|\Psi_{\rm VBS}^{(m,n)}\rangle
  =(-1)^n[1-{\cal O}(1/L)].
  \label{eqn:sign_relation}
\end{equation}
It turns out that $z_L$ changes its sign according to the number of
valence bonds at the boundary.  The value of $z_L$ for $m=n=1$ was
calculated by Totsuka and Suzuki~\cite{Totsuka-S}.  A similar relation
was also found by Bonesteel who discussed two-dimensional $S=1/2$ dimer
systems \cite{Bonesteel}.  In Eq.~(\ref{eqn:sign_relation}), the factor
$(-1)^n$ originates form the relative twist angle $2\pi$ at the
boundary, so that it does not depend on the detailed structure of the
twist operator (\ref{eqn:twist_op}). Therefore, the definition of $z_L$
is still meaningful even in cases with broken translational symmetry
($m\neq n$) which does not accord with the LSM argument.  Especially, in
the present definition, $z_L$ has a useful symmetry
$z_L\rightarrow(-1)^{2S} z_L$ under interchange of $m$ and $n$.

In order to calculate $z_L$ in more detail, we introduce the matrix
product (MP) formalism \cite{Fannes-N-W,Klumper-S-Z}. This is useful to
calculate the ground-state expectation values and correlation functions.
In this formalism, Eq.~(\ref{eqn:VBS}) is expressed as
\begin{equation}
 |\Psi_{\rm VBS}^{(m,n)}\rangle=
  \frac{1}{\sqrt{\cal N}}
  {\rm Tr}\,g_1^A\otimes g_2^B\otimes\cdots\otimes g_{L-1}^A\otimes g_L^B.
\end{equation}
These matrices for general integer $S$ case with $m=n$ are given in
Ref.~\cite{Totsuka-S}. Based on this result, the matrices for the
twisted VBS state $U^q|\Psi_{\rm VBS}^{(m,n)}\rangle$ are given by
\begin{eqnarray}
 \lefteqn{g_{q,j}^A(s,r)=(-1)^{n-s+1}{\rm e}^{{\rm i}(n-2s+2)q\pi/L}}
  \nonumber\\
 &&
  \times\sqrt{_m{\rm C}_{r-1} \,_n{\rm C}_{s-1}}
  (a_j^{\dag})^{m-r+s}(b_j^{\dag})^{n+r-s}|{\rm vac}\rangle_j,\\
 \lefteqn{g_{q,j}^B(r,s)=(-1)^{m-r+1}{\rm e}^{{\rm i}(m-2r+2)q\pi/L}}
  \nonumber\\
 &&
  \times\sqrt{_m{\rm C}_{r-1} \,_n{\rm C}_{s-1}}
  (a_j^{\dag})^{m-r+s}(b_j^{\dag})^{n+r-s}|{\rm vac}\rangle_j,
\end{eqnarray}
where $1\leq r\leq m+1$, $1\leq s\leq n+1$, and $|{\rm vac}\rangle_j$ is
the boson vacuum at the $j$-th site.  By introducing a transfer matrix
defined as
\begin{equation}
 G_q^{A}(r_1,r_2;s_1,s_2)\equiv
  g_{0,j}^{A\dag}(r_1,s_1)g_{q,j}^{A}(r_2,s_2),
\end{equation}
we obtain $z_L$ as
\begin{equation}
 z_L^{(1)}=(-1)^n
  \frac{{\rm Tr}\,[G_1^A G_1^B]^{L/2}}
  {{\rm Tr}\,[G_0^A G_0^B]^{L/2}}.
  \label{eqn:z_L_mp}
\end{equation}
Although the analytic form of $z_L$ is complicated in general, one can
evaluate numerical values of Eq.~(\ref{eqn:z_L_mp}), which will be shown
in Fig.~\ref{fig:z_L} below.  In the fully dimerized case
$(m,n)=(2S,0)$, $z_L$ is given by a simple form
$z_L^{(1)}=[\sum_{k=0}^{2S}{\rm e}^{{\rm i}(2S-2k)\pi/L}/(2S+1)]^{L/2}$.

One can show that $z_L$ behaves similarly even in open systems. For even
$L$ (see Fig.~\ref{fig:VBSpics}(b)), the VBS state of
Eq.~(\ref{eqn:VBS}) has $(n+1)^2$-fold degeneracy due to the isolated
bosons at both ends \cite{Oshikawa}.  Then, $z_L$ changes its sign
according to the number of isolated bosons at the $L$-th site,
\begin{equation}
 z_L^{(1)}=(-1)^n{\rm e}^{-{\rm i}n\pi/L}[1-{\cal O}(1/L)].
\label{eqn:z_L_open}
\end{equation}
Although $z_L$ is complex in this case, the imaginary part vanishes in
the large-$L$ limit, and the real part of $z_L$ behaves in the same way
as that of the periodic cases.  For open systems with odd $L$, the
number of isolated bosons at the $L$-th site is $m$, and the ground
state has $(m+1)(n+1)$-fold degeneracy.  Therefore, $z_L$ is simply
given by Eq.~(\ref{eqn:z_L_open}) multiplied by $(-1)^{2S}$.

\begin{figure}[t]
 \includegraphics[height=2.4in]{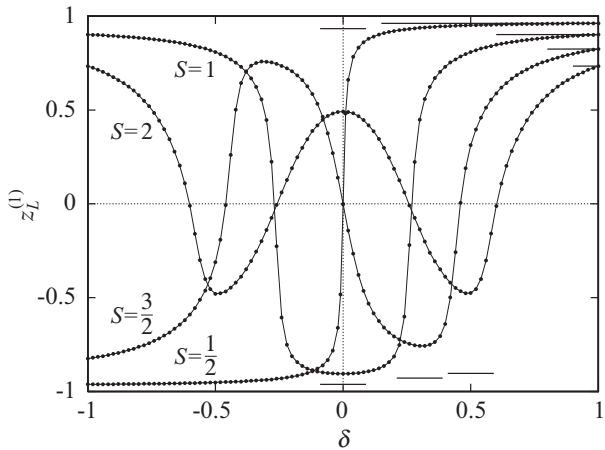}
 \vspace*{-1em}
 \caption{$\delta$ dependence of $z_L^{(1)}$ of the $L=64$ periodic BAHC
 with $S=\frac{1}{2}$, 1, $\frac{3}{2}$, and 2, obtained by the QMC
 calculation.  The horizontal lines indicate $z_L$ for the VBS states
 calculated by the MP method [Eq.~(\ref{eqn:z_L_mp})].}\label{fig:z_L}
\end{figure}

Next we consider $z_L$ in connection with a low energy effective
theory. According to Schulz's bosonization analysis \cite{Schulz}, the
Lagrangian density of the uniform Heisenberg spin chain is given by the
sine-Gordon model,
\begin{equation}
 {\cal L}=\frac{1}{2\pi K}\left[\nabla \phi(x,\tau)\right]^2
  -\frac{y_{\phi}}{2\pi\alpha^2}\cos[q\sqrt{2}\phi(x,\tau)],
  \label{eqn:SG_model}
\end{equation}
where $\tau$ is the imaginary time, $\alpha$ is a short range cut off,
and $K$ and $y_{\phi}$ are the parameters determined phenomenologically.
The phase field is related to that used in Ref.~\cite{Schulz} by
$\phi=2\sqrt{S}\psi_1$, and $q=1$ ($q=2$) for $S$ integer ($S$ half-odd
integer), where $q$ corresponds to degeneracy of gapped ground states.
In the gapped (gapless) region one has $y_{\phi}(l)\rightarrow\pm\infty$
($y_{\phi}(l)\rightarrow 0$) for $l\rightarrow\infty$ under
renormalization $\alpha\rightarrow{\rm e}^{l}\alpha$.  On the unstable
Gaussian fixed line [$y_{\phi}(0)=0$ with $Kq^2<4$], a second-order
``Gaussian transition'' takes place between the two gapped states.  In
this formalism, and the spin wave excitation created by $U$ corresponds
to the vertex operator $\exp({\rm i}\sqrt{2}\phi)$, so that $z_L$ for
$q(S-m)=$ integer is related to the ground-state expectation value of
the nonlinear term as
$z_L^{(q)}\propto\langle\cos(q\sqrt{2}\phi)\rangle$ and the three fixed
points $y_{\phi}=\pm\infty,0$ correspond to $z_{\infty}=\mp 1,0$,
respectively. Thus the Gaussian critical point can be identified by
observing $z_{L}^{(q)}=0$ \cite{Nakamura-V}.

\begin{figure}[t]
 \includegraphics[height=2.4in]{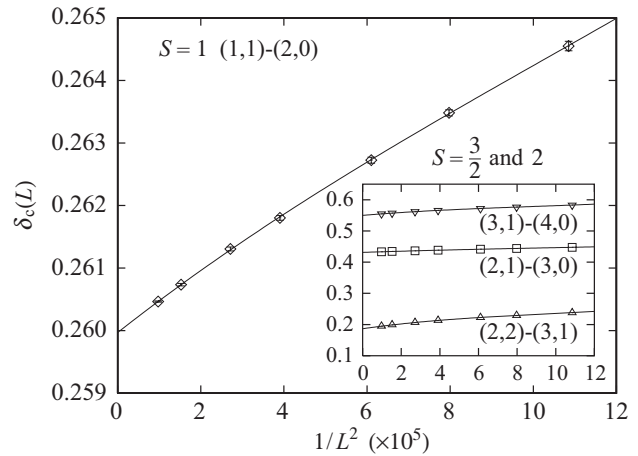}
 \vspace*{-1em}
 \caption{System size dependence of the successive dimerization
 transition points $\delta_{\rm c}(L)$ of the periodic BAHC with
 $S=\frac{1}{2}$, 1, $\frac{3}{2}$, and 2.  The extrapolation to the
 $L\rightarrow\infty$ limit is done by $\delta_{\rm c}(L)=\delta_{\rm
 c}(\infty)+A/L^2+B/L^4+C/L^6$ (solid lines).}\label{fig:z_L=0}
\end{figure}

In order to demonstrate the above argument, we consider the
bond-alternating Heisenberg chain (BAHC),
\begin{equation}
 {\cal H}=J\sum_{j=1}^L[1-\delta(-1)^{j}]\bm{S}_{j}\cdot\bm{S}_{j+1}.
  \label{eqn:BAHC}
\end{equation}
For this model, the VBS picture is considered to be realized
approximately: The configuration of the valence bonds $(m,n)$ changes
from $(0,2S)$ to $(2S,0)$ successively as $\delta$ is increased form
$-1$ to $1$, meaning the existence of $2S$ quantum phase transitions
\cite{Affleck-H,Oshikawa}. Since the translational symmetry is
explicitly broken, the effective model of the BAHC is given by
Eq.~(\ref{eqn:SG_model}) with $q=1$ for any $S$ \cite{Schulz}, and these
transitions are described as a Gaussian type.

In Fig.~\ref{fig:z_L}, we show $z_L$ (with $q=1$) of the $L=64$ periodic
BAHC with $S=\frac{1}{2}$, 1, $\frac{3}{2}$, and 2 as a function of
$\delta$.  For the calculation of $z_L$, we employ the QMC method with
the continuous-time loop algorithm \cite{Todo-K}.  We used the
multi-cluster variant of the loop algorithm.  The QMC steps are $10^3$
for thermalization, and $5\times 10^5$--$10^6$ for the measurement, and
the inverse temperature $\beta$ is taken large enough so that the value
of $z_L$ can be identified as that of the ground state.  The largest
inverse temperature used in this calculation is $\beta J = 128$.

As seen clearly in Fig.~\ref{fig:z_L}, the $\delta$ dependence of $z_L$
agrees qualitatively with the present interpretations based on the VBS
picture, which predicts $z_{\infty}=(-1)^n$ with a symmetry
$z_L\rightarrow(-1)^{2S} z_L$ for $\delta\leftrightarrow-\delta$
($m\leftrightarrow n$).  We have also calculated $z_L$ for open chains
with even and odd $L$'s, and confirmed that the results agree with our
predictions including Eq.~(\ref{eqn:z_L_open}).  In Fig.~\ref{fig:z_L},
we also present $z_L$ for the VBS states calculated by the MP method
[Eq.~(\ref{eqn:z_L_mp})] as horizontal lines.  The difference between
$z_L$ for the BAHC and that for the VBS states becomes larger as $S$
increases.  This means that the VBS picture becomes poor for the BAHC
with large $S$.

Next, we determine the critical point $\delta_{\rm c}$ by observing
$z_L=0$ with $L$ up to $320$ (Fig.~\ref{fig:z_L=0}).  In this
calculation, the inverse temperature is taken as $\beta J=L/2S$, being
assumed the Lorentz invariance.  Extrapolation to the
$L\rightarrow\infty$ limit has been done by the least-squares fitting by
assuming the asymptotic form as $\delta_{\rm c}(L)=\delta_{\rm
c}(\infty)+A/L^2+B/L^4+C/L^6$.  This polynomial with even powers of
$1/L$ is justified by the parity symmetry which ensures that $z_L$ is
real [Eq.~(\ref{eqn:z_L2})].  For $S=1$, we have obtained the critical
value for the transition between the $(1,1)$ and $(2,0)$ phases as
$\delta_{\rm c}=0.25997(3)$, where $(\,)$ denotes $2\sigma$.  This
result is consistent with the previous estimates: $\delta_{\rm
c}=0.2595(5)$ by the QMC calculation for the susceptibility
\cite{Kohno-T-H} and $0.2598$ by the level-crossing method
\cite{Kitazawa-N}.  Similarly, we identify the critical point of the
$S=3/2$ case [$(2,1)$-$(3,0)$] and those of the $S=2$ case
[$(2,2)$-$(3,1)$, $(3,1)$-$(4,0)$].  We obtain $\delta_{\rm
c}=0.43131(7)$, 0.1866(7), and 0.5500(1), respectively (see the inset of
Fig.~\ref{fig:z_L=0}).  They are also consistent with $\delta_{\rm
c}=0.4315$, 0.1830, and 0.5505 obtained by the level-crossing
method~\cite{Kitazawa-N2}, but much more accurate.

Here, we comment on the method used in Refs.~\cite{Kitazawa-N} and
\cite{Kitazawa-N2} proposed by Kitazawa who pointed out that the
Gaussian transition with $q=1$ can be identified by a level crossing of
excitation spectra under twisted boundary conditions \cite{Kitazawa}.
Since this method is also explained by the sine-Gordon theory
\cite{Kitazawa} and the VBS picture \cite{Kitazawa-N,Kitazawa-N2}, the
results are considered to be equivalent to ours.  Although finite-size
corrections in the level-crossing point in VBS states tend to be smaller
than those of the $z_L=0$ point, the application of the level-crossing
method to larger systems is difficult, because it relies on the exact
diagonalization to obtain excitation spectra which needs the whole
Hilbert space.  On the other hand, our approach is based only on the
ground state quantity, $z_L$, so that various numerical methods such as
the present QMC method can be employed.  This makes it possible to deal
with systems with enormous Hilbert space, such as large-$S$ and ladder
systems.  Note that the density matrix renormalization group method is
also suitable for our approach, since $z_L$ remains as a meaningful
order parameter even in open systems.  Furthermore, our method also has
an advantage to comprehend the physical picture as already discussed.

Finally, we discuss the relation between $z_L$ and the string order
parameter.  In Ref.~\cite{Oshikawa}, Eq.~(\ref{eqn:string}) for
$|\Psi_{\rm VBS}^{(m,n)}\rangle$ was calculated as ${\cal O}_{\rm
string}^z=[\frac{m+n+2}{2(m+2)}]^2 \delta_{n,{\rm odd}}$, where finite
and vanishing ${\cal O}_{\rm string}^z$ correspond to negative and
positive $z_L$, respectively.  However, this calculation is limited only
for $S$ integer, and the twist angle $\pi$ in Eq.~(\ref{eqn:string}) is
needed to be generalized for other cases \cite{Oshikawa}.  In addition,
it is difficult to determine accurate phase transition points by ${\cal
O}_{\rm string}^z$, because it does not change the sign.  Thus $z_L$
turns out to be more rational order parameter to describe VBS ground
states in a unified way.

In summary, we have introduced $z_L$ given as the ground-state
expectation value of the twist operator (\ref{eqn:def_z}) as an order
parameter to characterize various ground states in quantum spin
chains. Especially, $z_L$ changes the sign according to the
configuration of valence bonds.  This property enables us to determine
the critical point between different VBS states by observing $z_L=0$.
We have demonstrated this theory by using the QMC simulation for the
successive dimerization transitions of the BAHC, and determined the
phase boundaries with quite high accuracy.

M.N. thanks M. Oshikawa and J. Voit for useful discussions, and Y. Oi
Nakamura for encouragement.  S.T. acknowledges support of the Swiss
National Science Foundation.  The simulations were performed on SGI 2800
at the Supercomputer Center, Institute for Solid State Physics,
University of Tokyo.


\end{document}